# GLOBAL RECESSIONS AS A CASCADE PHENOMENON WITH HETEROGENEOUS, INTERACTING AGENTS

**Paul Ormerod, Volterra Consulting, London**

135c Sheen Lane

London SW14 8AE

UK

May 2008

pormerod@volterra.co.uk

*Abstract*

*This paper examines global recessions as a cascade phenomenon. In other words, how recessions arising within one or more countries might percolate across a network of connected economies. A heterogeneous agent based model is set up in which the agents are Western economies. A country has a probability of entering recession in any given year and one of emerging from it the next.*

*In addition, the heterogeneous agents each have a unique threshold propensity to import a recession from the agents most closely connected to them. The agents are connected on a small world topology, and an agent's neighbours at any time are either in (state 1) or out (state 0) of recession. If the weighted sum exceeds the threshold, the agent also goes into recession.*

*Annual real GDP growth for 17 Western countries 1871-2006 is used as the data set. The distribution of the number of countries in recession in any given year is exponential. The duration of recessions in the individual countries is also exponential. The model is calibrated against these facts and is able to replicate them. In addition it replicates the distribution of 'wait time' between recessions i.e. the number of years between them.*

*The network structure is essential for the heterogeneous agents to replicate the stylised facts. The country-specific probabilities of entering and emerging from recession by themselves give results very different to the actual data.*

1. **Introduction**

There is a very large literature within economics on business cycles within individual economies, going back at least as far as the work of Marx in the 19$^{th}$ century. It is fair to say that no genuine consensus has emerged within the profession of either the causes of cycles in general or recessions in particular. In a different area of economics, the last decade or so has seen considerable interest on information cascades and herding behaviour (for example, Banerjee (1992) and Bikchandani et.al. (1992)).

The aim of this paper is to examine global recessions as a cascade phenomenon. In other words, how recessions arising within one or more countries might percolate across a network of connected economies.

There is a variety of ways in which recessions might spread in this way. In 1932, for example, Keynes described how recessions could be spread by non-co-operative policies: 'Each nation, in an effort to improve its relative position, takes measures injurious to the absolute prosperity of its neighbors; and, since its example is not confined to itself, it suffers more from similar action by its neighbors than it gains by such action itself. Practically all the remedies popularly advocated to-day are of this internecine character. Competitive wage reductions, competitive tariffs, competitive liquidation of foreign assets, competitive currency deflations, competitive economy campaigns—all are of this beggar-my-neighbor description.'

Empirical evidence for a different mechanism is provided by, Eichengreen et.al. (1996). They use a panel of quarterly data for 20 industrial countries for the period 1959-1993 to show that currency crises and associated losses of real output spread more easily between countries which are closely tied by international trade linkages. Less formally, recessions might spread through the international percolation of expectations, of which the current sub-prime lending crisis is a potential example. Foster and Fleith (2002), for example, consider a nonlinear model in which agents exchange their opinions and information with each other. Their model yields multiple equilibria, or attracting distributions, that are persistent but subject to sudden large jumps

The above list is not meant to be exhaustive, nor is it my intention to offer a description of the exact mechanism by which recessions might cascade across countries. Rather, I consider evidence over a long time period of the distribution of the number of leading Western economies in recession in any given year, and examine the extent to which this can be explained by a model based on the principles of cascades.

Section 2 describes the data and the key empirical phenomena I require the model to be able to replicate. Section 3 sets out the theoretical model, and section 4 discusses the results.

## 2. The data and the stylised facts

Annual real GDP data for 17 countries 1870-1994 is taken from Maddison (1995). The 1995-2006 data is from the IMF database. Strictly speaking, the two sources are not exactly comparable since the Maddison data is in real Geary-Khamis dollars and the IMF in real domestic currency, but given that we are working with annual GDP growth, this is of little consequence.

The countries[1] are: Australia, Austria, Belgium, Canada, Denmark, Finland, France, Germany, Italy, Japan, Netherlands, New Zealand, Norway, Sweden, Switzerland, United Kingdom and United States.

The definition of a recession is the conventional one of a year in which the rate of growth of real GDP in a country is less than zero. Bordo and Helbing (2003), for example, discuss the advantages and disadvantages of NBER-based definitions of recessions, which consider a wider range of indicators. However, as they note, the two approaches give very similar results in terms of the timing of recessions. Further, and decisive in terms of the choice of definition of recession in this context, NBER-type business cycle dates do not exist for all the countries over the 1870-2006 period.

In terms of the total output of the 17 countries, the growth rate was less than zero in only 14 out of the total of 136 years in the data sample. For reference these are 1876, 1893, 1908, 1914, 1917, 1919, 1921, 1930-32, 1945-46, 1975 and 1982.

In no fewer than 40 out of the 136 years, no single country was in recession, and there is no example of a year in which all 17 were simultaneously in recession. The largest single number in any one year is 15 in 1931, closely followed by 14 in 1930. Of the 17 countries, 12 were in recession in 1914 and 1932, and 11 in 1917, 1918 and 1975.

The histogram of the number of countries in recession in any one year is set out below in Figure 1.

**Figure 1 here**

The null hypothesis that the data, with each frequency expressed as a percentage of the total, follow an exponential distribution with rate parameter of 0.26 is only rejected on a Kolmogorov-Smirnov test at a p-value of 0.529. (The p-value is maximised with this rate parameter). In other words, it is definitely not rejected at the conventional level of p = 0.05.

A second 'stylised fact' of interest is the distribution of the length of recessions. In total, there are 255 instances of recessions in the 17 capitalist economies over the 1871-2006 period. Of these around two-thirds, 164 to be exact, lasted only one year and 58 lasted 2

---

[1] In the Maddison data set, Swiss GDP data is only available but only from 1900 on an annual basis. And the data for Japan from 1885

years. Of the remainder, in 20 cases a country was in recession for three successive years, in 6 cases for 4 years, in 5 for 5 years, and there is one example each of recessions lasting 6 and 7 years respectively.

Given the small sample, with the longest lasting only 7 years, a formal statistical test for the distribution of the data is not all that helpful, but a non-linear least squares regression[2] of the frequency against the exponential of the length in years gives the following fit:

| *Duration: years* | 1 | 2 | 3 | 4 | 5 | 6 | 7 |
|---|---|---|---|---|---|---|---|
| Actual | 164 | 58 | 20 | 6 | 5 | 1 | 1 |
| Fitted | 164 | 58 | 20 | 7 | 3 | 1 | 0 |

The data is perhaps slightly more stretched than a simple exponential (Laherrere and Sornette (1998)), but the latter gives a very good fit to the data.

In terms of the distribution of the number of years between recessions, there are unavoidable problems with the data, relating to truncation. For each country, the years prior to the first recorded recession in the data set are ignored, as are the years to 2006 at the opposite end of the time period. So, for example, the last recorded recession in no fewer than 7 countries was in 1993, but for them the period 1993-2006 is ignored. For recessions of longer duration than a single year, the period between recessions is counted as one year for the second and each subsequent year.
**Figure 2 here**

The theoretical model is therefore to be validated against:

- its ability to generate the relevant exponential distributions for the number of countries in recession in any one year
- its ability to generate the distribution of the length of recessions
- the distribution of the number of years between recessions

### 3. The cascade model

The model is populated by 17 countries, the sizes of which correspond to the size of the actual economies in dollar terms in 1955. The year 1955 is chosen because it is close to the mid-point of the sample period but is not subject to the distortions of either the Great Depression of the 1930s or of World War Two and its aftermath.

---

[2] Carried out with the command 'nls' in the package S-Plus

The model proceeds in a series of steps. In the first step, each country goes into recession with probability $\pi$. The mechanism by which an individual country enters recession of its own accord, as it were, is not specified, but the key feature is that it is stochastic, a feature, for example, of real business cycle theory.

In step 2 of the solution, those which were in recession in step 1 recover with probability $\rho$. The remainder enter recession with probability $\pi$. Both $\pi$ and $\rho$ are drawn from a uniform distribution at each step, so there is heterogeneity both across countries at a point in time and within a country across time in terms of the propensity to enter into or recover from recession. The model proceeds over a total of 136 steps, the length of the sample period 1871-2006.

Each economy is connected to a particular number of neighbours and recessions are propagated across the connections. Each economy is assigned a threshold chosen at random from a uniform distribution on $[\tau, 1]$. The neighbours are assigned a value of 0 if not in recession and 1 if they are in recession. The economy goes into recession if the weighted sum of the neighbours in the period is greater than $\tau$, where the weights are the sizes of the economies. So there is heterogeneity amongst the economies in terms of their propensity to 'import' recession at any given point in time,

There are therefore two sources of recession, the stochastic element $\pi$ which is particular to each individual economy, and the propagation mechanism across the network of economies.

Evidence on the nature of the relevant topology of the network of connections is sparse. Serrano and Boguna (2003), for example, show that the world trade network is approximated by a small world topology. Small world networks (Watts and Strogatz (1998)) consist of a network of vertices whose topology is that of a regular lattice, with the addition of a low density $\varphi$ of connections between randomly-chosen pairs of vertices. This latter can be either an addition to the connections on a regular lattice or can replace the relevant number of such connections.

The initial investigations with the model were carried out with a small world network, and the results proved to be very satisfactory. There are two parameters to select, the number of immediate neighbours of a country, k, and the probability of a connection between any randomly-chosen pairs of countries, $\mu$. In the calibration of the model, such connections replace connections with immediate neighbours so that the number of neighbours remains the same for each agent. The word 'neighbour', to emphasise, does not mean geographic neighbour, but the countries which are closest to any given country in terms of their influence on whether or not it goes into recession.

We do not know at any point in time which countries exercise the most influence on any other. We therefore obtain 5,000 separate solutions of the model, each over 136 steps. In each solution, the exact structure of the small world network is drawn afresh, with parameters k and $\mu$.

There are five parameters to choose in calibrating the model, $\pi$, $\rho$, $\tau$, k and $\mu$. After experimentation, the following were selected: $\pi$ is drawn from a uniform distribution on [0.01, 0.11], $\rho$ is drawn from a uniform distribution on [0.76, 1.0], $\tau = 0.1$, $k = 2$, $\mu = 0.08$.

The results are to a considerable extent robust to the choice of the re-wiring parameter, $\mu$. The graph theoretic properties of the small world network used in the model show why. Figure 3 plots the average path length of a small world graph with 17 vertices. The blue line plots the path length with each vertex connected to one neighbour either side, for different values of the rewiring parameter, and the red line shows the same but with two neighbours either side. The results show the average over 500 separate realisations of the graph.
**Figure 3 here**

.

The average path length is the average number of steps along the shortest paths for all possible pairs of network vertices. A short average path length facilitates the transfer of information and/or influence. Figure 1 shows that for $k = 2$, the average path length does not vary much with the probability of rewiring, which establishes direct, longer-range connections in the graph. Even with no rewiring, the average path length is only 2.5.

As Figure 1 shows, the distribution of the number of countries in recession in any given year (n) is slightly erratic when n = 1 and 2, though overall the distribution is exponential. The model-generated results averaged over 5,000 simulations do not replicate this eccentricity, because the process of averaging smooths out the fluctuations contained within individual solutions. However, the null hypothesis that the actual data and the model-generated results the two are the same is only rejected on a Kolmogorov-Smirnov test at a p-value of 0.972. The simple correlation between the two is 0.928

The model does produce a very small percentage of periods (1.05) when there is a truly global recession in which all 17 countries are in recession in the same year.

In terms of the duration of recessions generated by the model, 99.4 per cent of all recessions are between 1 and 7 years. The very large number of observations (1,169,451 recessions) generated by the model leads to a longer 'tail' in terms of duration than in the actual data. The longest is in fact 30 years, although only 0.15 per cent of recessions are longer than 10 years.

But the proportion of recessions by duration is very similar in the model output and the actual data.

|  | Duration | | | | | | |
| --- | --- | --- | --- | --- | --- | --- | --- |
|  | 1 | 2 | 3 | 4 | 5 | 6 | 7 |
| Actual | 0.643 | 0.227 | 0.078 | 0.024 | 0.020 | 0.004 | 0.004 |
| Model | 0.626 | 0.217 | 0.085 | 0.036 | 0.017 | 0.009 | 0.004 |

A non-linear least squares regression of the actual data on the exponential of a constant and the number of years gives a parameter of -1.043 with a standard error of 0.018, and for the model output ( using the full span of durations up to 30 years ) it is -1.023 with a standard error of 0.010.

The correlation between the actual wait time between recessions and the model generated one is 0.998. The null hypothesis that the actual data and the model-generated results of the two for wait times of between 1 and 25 years are the same is only rejected on a Kolmogorov-Smirnov test at a p-value of 0.476and for 1 and 31 years at 0.256. Over the data set as a whole the null hypothesis is rejected at $p < 0.05$, because of the very long tail which the model results contain. In the actual data, there is only one observation with a wait time between recessions of more than 31 years, which is 0.26 per cent of the total number of observations. The very large number of runs of the model generate wait times of up to 124 years, with 1.26 per cent of the total number of observations being above 31 years. So over the bulk of the data, the model accurately replicates the data.

This simple model is therefore able to replicate three key empirical features of recessions.

It is useful to examine the role which the network plays in generating these results. We can do so by retaining the values $\pi$ in [0.01, 0.11] and $\rho$ in [0.76, 1.0], the probabilities of any individual country going into and coming out of recession in the absence of any influence from its neighbours, and by removing the network. In other words, by setting $\nu$ and $\mu$ equal to zero.

Again with 5000 separate solutions of the model, the output with no percolation across a network is very different from the actual data. Of the total number of model-generated recessions under these assumptions, 83 per cent last only 1 year and 14 per cent last 2 years, compared to the actual 64 and 23 per cent respectively. In terms of the number of countries in recession in any one period, the maximum number generated by the model is only 4, compared to 15 in the actual data.

So the percolation aspect of the network is decisive in enabling the model to generate results which give a good approximation to reality.

### 4. Conclusion

Two features of recessions under capitalism, using a data base of annual real GDP growth 1871-2006 in 17 Western economies, are, first, that the number of countries in recession in any given year follows an exponential distribution. Second, that the duration of recessions within individual countries also follows an exponential distribution.

A model of heterogeneous, interacting agent is able to replicate these two key facts. The agents are countries, which have a probability of both entering and emerging from recession. In addition, they are connected on a small world topology. The neighbours of any given agent are assigned the value 1 if they are in recession and 0 if not. If the weighted sum of neighbours exceeds a random, heterogeneous threshold, the agent also enters recession. The role of the network is decisive in the ability of the model to replicate the key facts on recession.

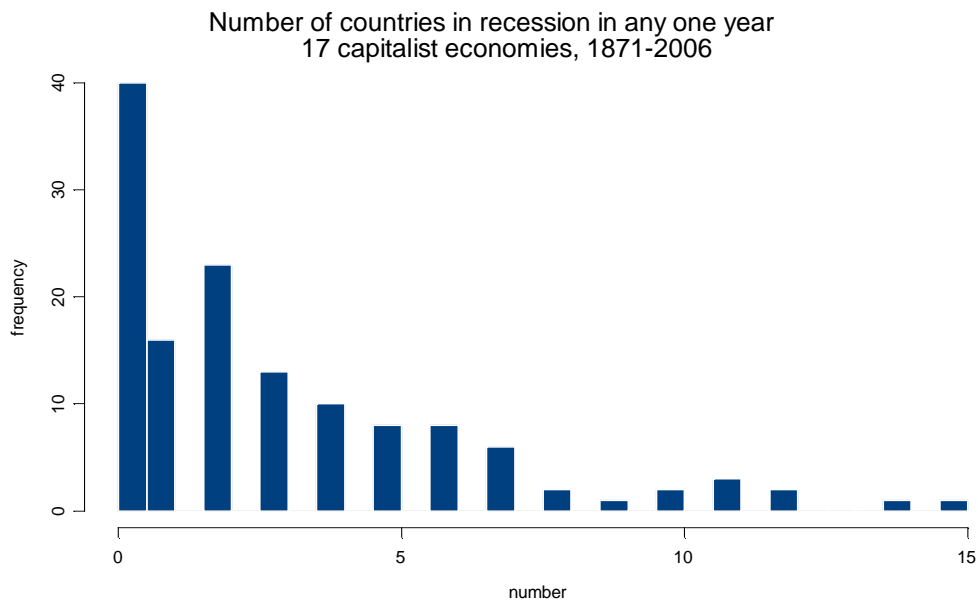

Number of countries in recession in any one year
17 capitalist economies, 1871-2006

**Figure 1** *Number of countries in recession in any one year, 17 countries, 1871-2006*

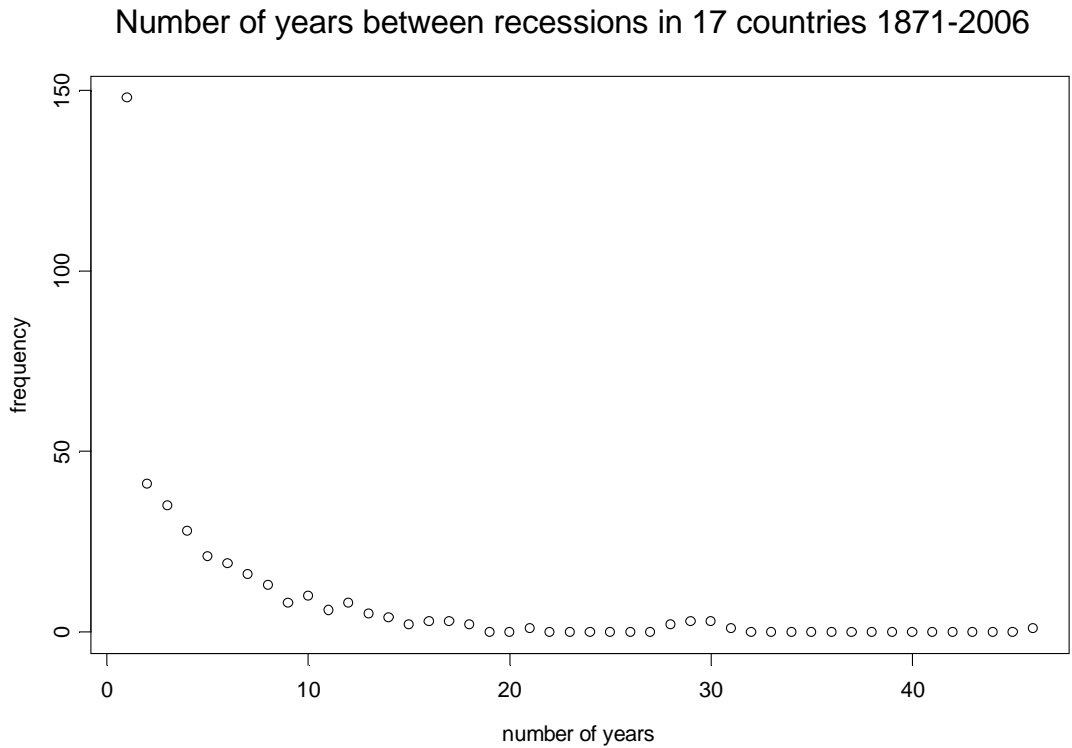

**Figure 2** *Distribution of number of years between recessions, 17 Western economies, 1871-2006; truncated data. The most frequent observation is 1 year i.e. observed when recessions last more than a single year*

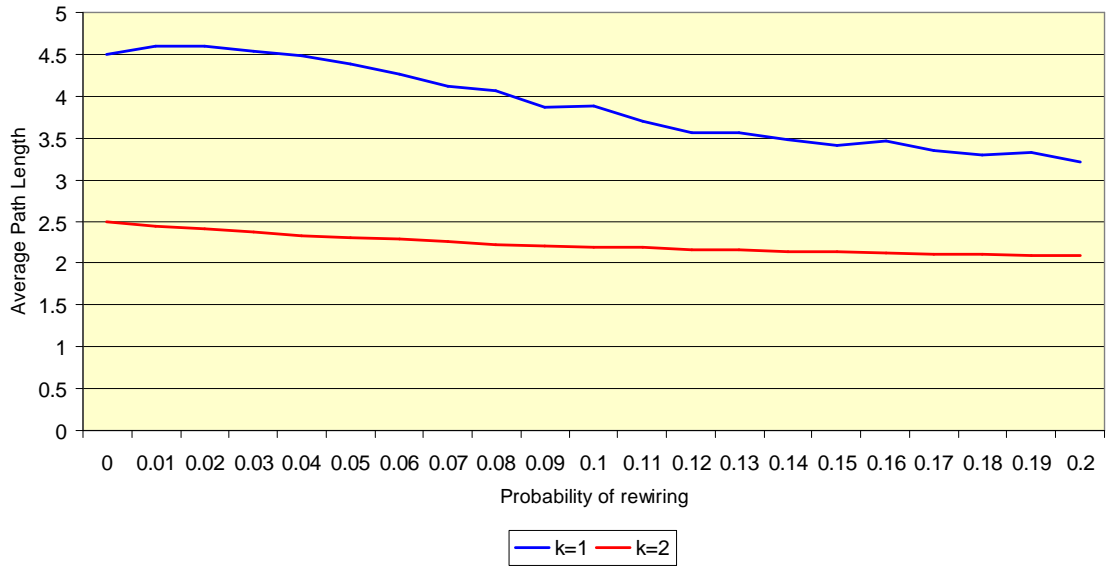

**Figure 3** *Average path length of small world network with 17 vertices, average over 500 separate realisations; rewiring replaces immediate neighbour connections*